\documentstyle[aps,prl]{revtex}         

\begin{document}

\draft

\title{The beta function of the multichannel Kondo model}
\author{Kurt Fischer}
\address{
Nagaosa group,
Department of Applied Physics,
University of Tokyo,
Tokyo 113-8656,
Japan}

\date{November 13}
\maketitle

\begin{abstract}
The beta function of the multichannel Kondo model is calculated exactly
in the limit of large spin $N$ and channel number $M=\gamma N$, with constant
$\gamma$. 
There are no corrections in any finite order of $1/N$.
One zero is found at a finite coupling strength, showing directly the 
Non--Fermi liquid behavior of the model.
This renormalization group flow allows to introduce a variational principle for
the entropy, to obtain the low temperature thermodynamics.
Such in particular the low temperature thermodynamics of the 
non--crossing approximation to the Kondo model becomes accessible.

\end{abstract}


Although the concept of a renormalization group flow in quantum field theory
has for a long time been established, the existence of a nontrivial fixed 
point can be shown usually~\cite{review-Wilson} only if it occurs near the 
trivial one of the noninteracting theory.
It is therefore perhaps worth mentioning that the well studied multichannel 
Kondo model is such a rare case, where the renormalization group flow can be 
obtained to all orders in the coupling constant, beyond the perturbative 
regime.

This model describes $M$ identical structureless conduction 
bands of width $D$, of fermions with an $N$ fold degenerate spin degree of 
freedom, which scatter via an spin--exchange interaction of strength $g/N$ at 
one localized impurity.
It is known from a physical argument~\cite{Nozieres/Blandin} and a $1/M$ 
expansion~\cite{Nozieres/Blandin,Gan/Andrei/Coleman-Non-Fermi} that the
renormalization group flow of this model should have at least one low 
temperature stable fixed point at a finite scattering strength, in contrast to
the one channel case~\cite{review-Wilson} where it is at infinite coupling. 
The critical behavior at low temperatures has been calculated for some 
quantities~\cite{Gan/Andrei/Coleman-Non-Fermi,Affleck/Ludwig-Non-Fermi,Cox/Ruckenstein,Jerez/Andrei/Zarand-Non-Fermi}, 
showing that the system is a Non--Fermi liquid for $M,N\geq 2$, and the 
critical exponents being a function of $\gamma = M/N$ only.

In this work, the result of 
Refs.~\onlinecite{Gan/Andrei/Coleman-Non-Fermi,Cox/Ruckenstein} is generalized
and the $\beta$ function is calculated in the limit of large $N$ with 
$\gamma= M/N$ kept constant.
There will be exactly one intermediate coupling fixed point 
$g_c\approx \gamma^{-1}$, and $\beta^\prime(g_c)$ is evaluated to be 
$1/(1+\gamma)$.
It is shown explicitly how the renormalization group flow depends on the 
cutoff scheme between its fixed points. 
There will be no corrections to the result in any finite order of $1/N$.
A variational principle for the low temperature entropy is introduced,
generalizing the one for $T=0$ in Ref.~\cite{Affleck/Ludwig-PRL-Non-Fermi}, 
and leading to the low temperature thermodynamics.

The multichannel Kondo Hamiltonian is given by
\begin{equation}\label{hamiltonian}
  H = \sum_{p,m,n} \epsilon_p c_{pmn}^+ c_{pmn}^{\phantom{+}} -
  \frac{g}{N} \sum_{p,q \atop m,n,n^\prime} 
  f_{n^\prime}^+  c_{qmn^\prime}^{\phantom{+}} c_{pmn}^+ f_n^{\phantom{+}} .
\end{equation}
Here $c_{pmn}^+$ creates a conduction electron with spin $n=1,\dots,N$ and 
scattering channel $m=1,\dots,M$, momentum $p$ and energy $\epsilon_p$.
For the density of states, a sharp cutoff at $\pm D$ is used.
The fermionic operator $f_n^+$ describes the magnetic impurity configuration 
$| n \rangle = f_n^+ | \text{vac} \rangle$.
The impurity side is supposed to be singly occupied: 
$\sum_n f_n^+ f_n \equiv 1$.
The Boltzmann constant and the density of states at the Fermi surface 
$\epsilon_p=0$ are set to unity.
For $g>0$, the interaction is antiferromagnetic. 

The diagrammatic technique for this model is standard~\cite{review-Bickers}.
Each interaction vertex in the scattering channel $m$ 
is represented by a bare pseudo propagator $S_m^{(0)}(z)\equiv g$ without 
dynamics, which is for short called a propagator.
The magnetic configurations have the bare propagators $R_n^{(0)}(z) = 1/z$.

The full propagators determine completely the physics of $H$, because every 
observable can be obtained from their Luttinger--Ward 
functional by coupling to an auxiliary field~\cite{review-Bickers,KuramotoI}.
To calculate the $\beta$ function of $H$, it is hence enough to show 
that in the low energy limit $z,T \ll D$, the propagators scale as functions 
of $z,T$ with the Kondo temperature $T_K$, which will be calculated, depending 
on the cutoff scheme, leading to a renormalization group flow, with a $\beta$
function defined as
\begin{equation}\label{beta-fct-def}
\partial_{\ln D} T_K = - \beta(g) \partial_g T_K .
\end{equation}
It was observed for the related multichannel 
Anderson model~\cite{Cox/Ruckenstein}
that the zero-temperature critical exponents of certain observables allow for 
a controlled expansion in the limit of large $M,N$ with fixed $\gamma$.
That limit will be introduced here as well.
It amounts to summing all diagrams with non-crossing conduction electron lines 
(NCA~\cite{Cox/Ruckenstein,review-Bickers}), because the $1/N$-order of a 
diagram for a propagator is not changed, if a self--energy of second order is 
eliminated, and just for these diagrams this procedure stops if the zero order 
contribution is reached.
Corrections are of relative order $1/N^2$ or smaller~\cite{review-Bickers}; 
these are dealt with later.

The Dyson equations in this limit are obtained by keeping the self--energies 
of second order in the Luttinger-Ward functional but inserting the dressed 
propagators~\cite{review-Bickers}
\begin{eqnarray}\label{R-S}
S^{-1}(z) = g^{-1} +  \int_{-D}^D f(\epsilon) R(z+\epsilon) d\epsilon 
\nonumber \\
R^{-1}(z) = z + E_0 
            + \gamma \int_{-D}^D f(\epsilon) S(z+\epsilon) d\epsilon 
\end{eqnarray}
where $f$ is the Fermi function, the indices $m,n$ have been dropped, and the
ground state energy $E_0$ has been subtracted.
Scaling for $R$ and $S$ will be established first for frequencies 
$T \ll -z  \ll D$ and then generalized to all $T,|z| \ll D$.
To this end, Eqs.~(\ref{R-S}) are differentiated with respect to $z$, 
neglecting terms $\propto 1/D$ and $O(\exp[-D/T])$,
\begin{eqnarray}\label{R-Integro-Deq}
\partial_z S^{-1}(z) &=&  \int_{-\infty}^\infty d\epsilon\; 
    \delta_T(z-\epsilon) R(\epsilon) \\
\partial_z R^{-1}(z) &=&  1 -  \gamma S(z-D) 
    + \gamma \int_{-\infty}^\infty d\epsilon\; 
    \delta_T(z-\epsilon) S(\epsilon) \nonumber 
\end{eqnarray}
where $\delta_T(z) = -\partial_z f(z)$.
For $T \ll -z $, Eqs.~(\ref{R-Integro-Deq}) can be converted into the 
temperature independent differential equations by neglecting terms of the 
order $O(T/z)$,
\begin{eqnarray}\label{R-differential-eq}
\partial_z S^{-1}(z) &=& R(z) \\
\partial_z R^{-1}(z) &=&  1 - \gamma S(z-D) + \gamma S(z) . \nonumber
\end{eqnarray} 
Because Eqs.~(\ref{R-Integro-Deq}),(\ref{R-differential-eq}) are independent 
of $g$, its value will be fixed implicitly by imposing the boundary condition 
\begin{equation}\label{S-boundary-condition}
S(-D) \stackrel{\text{def}}{=} \hat{g} .
\end{equation}
$\hat{g}$ will play the role of a dressed coupling, and coincides in the limit 
of small coupling with $g$ because of Eqs.~(\ref{R-S}).
This is the analogue of choosing the renormalization point in the usual, 
perturbative renormalization scheme. 
The second boundary condition is $R^{-1}(z=T=0) = 0$, because $R^{-1}$ 
vanishes at the $N\to\infty$ ground state energy $E_0$ which was fixed in 
Eqs.~(\ref{R-S}) at $z=0$.
The two boundary conditions determine the solutions 
of Eqs.~(\ref{R-Integro-Deq}),(\ref{R-differential-eq}).

Now choose $A$ with $T \ll A \ll D$. 
The function 
\begin{eqnarray}\label{integral}
I(z) &=& (\gamma^{-1} - \hat{g} ) ( S^{-1}(z) - \hat{g}^{-1} ) \\
&& + \ln ( S^{-1}(z)\hat{g} ) 
- \gamma^{-1} \ln ( R^{-1}(z)/ R(-D) ) \nonumber
\end{eqnarray}
is because of 
$\partial_z I(z) = R(z) ( S(z-D) - \hat{g} ) $
and Eq.~(\ref{S-boundary-condition}) an integral for 
Eqs.~(\ref{R-differential-eq}) for $T \ll -z < A$ if terms of the order $1/D$ 
are neglected.
Therefore in this frequency region the boundary condition 
Eq.~(\ref{S-boundary-condition}) can be replaced by the condition 
$I \equiv I(-A)$.
The propagators $R,S$ determine the single particle excitation spectrum of $H$ 
in the limit $N\to\infty$~\cite{review-Bickers}.
Hence the solutions of Eqs.~(\ref{R-differential-eq}) are real and have no 
singularity below the ground state energy corresponding to $z=0$.
Therefore and because the solutions of Eqs.~(\ref{R-differential-eq}) 
depend for $z< 0$ smoothly on the original boundary condition 
Eq.~(\ref{S-boundary-condition}), the quantity 
$\hat{D} \stackrel{\text{def}}{=} -R^{-1}(-D)\exp[-\gamma I(-A)]$
is a smooth, real function of $g$.
The rescaling $z\to z/D$, $R^{-1} \to R^{-1}/D$ in Eqs.~(\ref{R-S}) shows
that $\hat{D}/D$ is a function of $\hat{g}$ only.
If the Kondo temperature is now defined as
\begin{equation}\label{T-Kondo}
T_K = \hat{D} \hat{g}^\gamma | \gamma^{-1} - \hat{g} |^{-(\gamma+1)} 
      \exp[ \gamma -1/\hat{g} ] ,
\end{equation}
then after the rescaling $u = z/T_K$ and 
\begin{eqnarray}\label{rescaling}
r(u) &=& ( \gamma^{-1}- \hat{g} ) T_K R(z)  \\
s(u) &=& ( \gamma^{-1}- \hat{g} )^{-1} S(z) \nonumber 
\end{eqnarray}
the boundary condition $I\equiv I(-A)$ gives with Eq.~(\ref{integral}) the 
scale invariant integral 
\begin{equation}\label{scale-inv-boundary-cond}
\left[ -r^{-1}(u) \right]^{1/ \gamma} = s^{-1}(u) \exp [ s^{-1}(u)] .
\end{equation}
Applying now Eq.~(\ref{rescaling}) to Eqs.~(\ref{R-differential-eq}), it is 
evident that it scales with $T_K$ for $|z| \ll D$, because its integral
 Eq.~(\ref{scale-inv-boundary-cond}) does.
However, Eqs.~(\ref{R-differential-eq}) is the high frequency limit 
$-z \gg T$ of Eqs.~(\ref{R-Integro-Deq}). 
If therefore additionally $T\to T/T_K$, its solution also scales with $T_K$ if 
$T,|z| \ll D$; hence the solution of Eqs.~(\ref{R-S}).
This shows that there is a renormalization group flow, with a low
temperature fixed point according to Eq.~(\ref{T-Kondo}) at 
$\hat{g}_c =  \gamma^{-1}$.
The value $g(\hat{g}_c)$ depends via Eq.~(\ref{S-boundary-condition}) on the 
particular cutoff-scheme.
In terms of the $\beta$ function, defined by Eq.~(\ref{beta-fct-def}), 
the main result reads as 
\begin{equation}\label{beta-fct-result}
\beta( \hat{g}) = -  \frac{ \hat{g}^2 (\gamma^{-1} -\hat{g} ) }
{ \gamma^{-1} + \hat{g}^2 + \alpha(\hat{g}) \hat{g}^2 (\gamma^{-1} -\hat{g} ) }
\end{equation} 
where $\alpha(\hat{g}) = \partial_{\hat{g}} \ln \hat{D}/D$ is a smooth, 
non universal function of $\hat{g}$.
This result is still valid when the coupling $g_c$ becomes large.
The then arising nonuniversal terms of the $\beta$ function can be entirely 
expressed in terms of $\hat{D}$ and $\hat{g}$. 
Two limiting cases are considered in the sequel.
For small $\gamma^{-1}$, assuming $g < g_c$, the expansion in $\gamma^{-1}$ is
possible,
\[
\beta(g) = - g^2 + \gamma g^3 + \gamma g^4 - \gamma^2 g^5 + O(\gamma^{-4}) 
         + \text{nonuniversal terms} 
\]
and coincides~\cite{Gan/Andrei/Coleman-Non-Fermi} with the large $M$ 
expansion for $N=2$.
In the other limit $g \approx g_c$, it is customary to use the 
running coupling constant 
$g(T)$, defined by $T \partial_T g(T) = \beta(g(T))$ with the initial 
condition $g(D)=g$.
Its low temperature form is determined by the universal number
$\beta^\prime(g_c) = 1/(\gamma + 1)$, as
\begin{equation}\label{running-g}
| g_c - g(T) | \propto T^{1/(1+\gamma)} .
\end{equation}
Next, the corrections at finite $1/N$ are considered.
The Luttinger--Ward functional is expanded in $1/N$, yielding diagrams for 
products of terms $R(z) \Sigma^{\nu}(R(z),S(z))$ with $\nu > 2$,
where $\Sigma^{(\nu)}$ is the $\nu$th order self--energy of the magnetic 
configuration propagator.
In each such diagram, the number of propagators $R$ equals those of $S$ and the
number of integrations $\int_{-D}^D f(\epsilon) d\epsilon$ over the energies of
the conduction electrons. 
After the rescaling of $T$, $z$, $\epsilon$, $R$, and $S$ with $T_K$, in the 
low energy limit $z,T \ll D$ as in Eq.~(\ref{rescaling}), all factors $T_K$ 
and $\gamma^{-1}- \hat{g}$ are canceling each other.
Hence the Luttinger-Ward functional still scales with $T_K$ at low energies, 
so that there are no corrections to the $\beta$ function at any finite order
in $1/N$.

In order to extract the low temperature thermodynamics of the model, 
consider its entropy $S$, first at small coupling $g \ll 1$ and temperatures 
$T_K < T \ll D$.
Then perturbation theory in $g$ can be applied, for each order in 
$T/D$~\cite{Kondo-Non-Fermi},
\begin{equation}\label{entropy-low-T-g}
S(T,g) = S_0 (g,T) + S_1(g,T) T/D  + O(T^2/D^2) .
\end{equation}
With Eqs.~(\ref{beta-fct-result}),(\ref{running-g}) this can be 
renormalized in terms of the running coupling constant $g(T)$, to give near 
$g_c$ 
\begin{equation}\label{entropy-running-g}
S = S_0 (g(T)) + \lambda_1 |g_c -g(T)|^{1+\gamma}  + \dots ,
\end{equation}
where $\lambda_1$ is a constant.
Now, if $T$ is lowered the entropy will decrease.
But because of the renormalization group flow, $S$ will decrease also
if $g(T) \to g_c$. 
Therefore $S$ has a minimum at $g=g_c$.
This is the generalization of the variational principle for the $T=0$ 
entropy~\cite{Affleck/Ludwig-PRL-Non-Fermi}.

There are two cases to be distinguished:
In the case $\gamma > 1$ it can be assumed that perturbation theory is valid 
for all $g\leq g_c$.
Then $S_0$ is dominating, because applying the variational principle to 
$S\approx S_0$, it can be expanded around its minimum $g_c$ to second order,
so that with Eq.~(\ref{running-g}) follows ($\lambda_2$ being a constant)
\[
S_0 = S_0(T=0) +  \lambda_2 T^ {2/(1+\gamma)} + \dots \quad  .
\]
This justifies the assumption that $S_0$ dominates the second term $\propto T$
in Eq.~(\ref{entropy-running-g}).
For $\gamma < 1$, the situation is opposite, giving for $\gamma\neq 1$ the 
specific heat 
\begin{equation}\label{entropy-low-T}
C(T)  \propto T^{\min \{ 2/(1+\gamma), 1 \} } +  \dots 
\end{equation}
In Ref.~\onlinecite{Parcollet/Georges-Non-Fermi}, this law
was obtained by a large $N$ expansion of Hamiltonian~(\ref{hamiltonian}) 
but with the different constraint $\sum_n f_n^+f_n^{\phantom{+}} = q_0 N$ 
with constant $q_0$, instead of the constraint 
$\sum_n f_n^+f_n^{\phantom{+}} =1$ used here.
Although those results cannot be expanded in $q_0 = 1/N$, the low temperature
thermodynamics of the two models coincide.
The reason is that the variational principle introduced above, works quite 
generally if the model has a renormalization group flow with a finite, low 
temperature fixed point $g_c$. 
The low temperature thermodynamics depends then only on $\beta^\prime(g_c)$, 
which coincides with the value obtained here.

To conclude, in this work the renormalization group flow and the $\beta$
function for the multichannel Kondo model have been derived, by a large $N$ 
expansion to all orders in $1/N$.
The dependence of the $\beta$ function on the cutoff, could be expressed in
terms of two smooth functions of the coupling constant. 
In contrast to the usual perturbative approaches to scaling, 
this method remains valid even if the running coupling constant becomes large. 
The zero temperature variational principle of 
Ref.~\onlinecite{Affleck/Ludwig-PRL-Non-Fermi} was generalized to the leading
low temperature expansion of the entropy, which was shown to have a minimum at 
the low temperature fixed point of the coupling constant flow.
This in turn yielded the low temperature thermodynamics of the model, 
to all orders in $1/N$.
This means in particular that  Eq.~(\ref{entropy-low-T}) describes the low 
temperature thermodynamics of the non-crossing approximation to the $SU(N)$ 
Kondo model.

Finally, it is a  pleasure for me to acknowledge the persistent support of 
Prof. N. Nagaosa.
The author kindly expresses his gratitude for the financial support by the 
Japan Society for the Promotion of Science.

\end{document}